\documentclass[prd,amssymb,10pt]{revtex4}
\usepackage{amsmath,amsfonts,amssymb}
\usepackage{verbatim,graphicx,float}

\begin{document}

\title{Smoothed quantum fluctuations and CMB observations}

\author{Jakub Mielczarek}
\email{jakub.mielczarek@uj.edu.pl}
\affiliation{\it Astronomical Observatory, Jagiellonian University, 30-244
Krak\'ow, Orla 171, Poland}

\author{Micha{\l} Kamionka}
\email{kamionka@astro.ini.wroc.pl}
\affiliation{ Astronomical Institute, Wroc{\l}aw University,
51-622 Wroc{\l}aw, Kopernika 11, Poland}

%\date{\today}

\begin{abstract}
In this paper we investigate power spectrum of a smoothed scalar field.
The smoothing leads to the regularisation of the UV divergences 
and can be related with the internal structure of the considered 
field or the space itself. We apply procedure of smoothing to the quantum 
fluctuations generated during the phase of cosmic inflation. 
We study whether this effect can be probed observationally
and conclude that the modifications of the power spectrum due to 
the smoothing on the Planck scale are negligible and far beyond the observational
abilities. Subsequently we investigate whether smoothing in any other form can be probed 
observationally. We introduce phenomenological smoothing factor $e^{-k^2\sigma^2}$ 
to the inflationary spectrum and investigate its effects on the 
spectrum of CMB anisotropies and polarisation. We show that smoothing 
can lead to suppression of high multipoles in the spectrum of the CMB. 
Based on five years observations of WMAP satellite we 
indicate that the present scale of high multipoles suppression is constrained by 
$\sigma < 2.86$ Mpc (95$\%$ CL). This corresponds to the constraint
$\sigma < 100 \ \mu$m at the end of inflation. Despite this value is far 
above the Planck scale, the other processes of smoothing can be possibly
studied with this constraint, e.g. diffusion or decoherence of primordial 
perturbations.
\end{abstract}

\maketitle

\section{Introduction} \label{sec:intro}

The ultraviolet (UV) divergences of a quantum filed theory are related to the short distance 
behaviour of the vacuum expectation values of product of the field operators. In case of 
the free scalar field in Minkowski background, the correlator is given by   
\begin{equation}
\langle 0|\hat{\phi}({\bf x},t)\hat{\phi}({\bf y},t) |0 \rangle =\frac{1}{4\pi^2} \frac{1}{({\bf x-y})^2}. \label{CM}
\end{equation}
In the limit $({\bf x-y})^2\rightarrow 0$ the correlator diverges leading to the bad UV behaviour. The one possible way to 
remove UV divergences is to introduce a cut-off on some sufficiently high energy scales. In this approach 
momentum integrations have to be limited to the scale of cut-off $\Lambda_{\text{cut-off}}$. Then field theoretical 
description of the field is effective and valid up to the scale of cut-off $\Lambda_{\text{cut-off}}$, where 
a new physical description has to be introduced. Another way to introduce cut-off is to perform averaging 
of the field on the sufficiently small distances. Then we introduce a new field $\hat{\phi}_{\Sigma}({\bf x},t)$ which 
is averaged version of the field $\hat{\phi}({\bf x},t)$ with some averaging function $\Sigma({\bf x})$ which fulfils
the normalisation condition $\int_{V} d^3 {\bf x}  \sqrt{q} \Sigma({\bf x}) = 1$. The $q$ is determinant of the spatial 
part of the metric. In our considerations we take the FRW metric, then $q=a^6$ where $a$ is the scale factor. 
It turns out that vacuum expectation values of product of the new fields $\hat{\phi}_{\Sigma}({\bf x},t)$ becomes 
finite for $({\bf x-y})^2\rightarrow 0$. Therefore the UV divergence is removed. In the model considered in this 
paper we assume smoothing Gaussian function in the form
\begin{equation}
\Sigma({\bf x})=\frac{1}{(2\pi\sigma^2)^{3/2}} e^{-\frac{|{\bf x}|^2a^2}{2\sigma^2}}. \label{Gauss}
\end{equation}
The parameter of this function is dispersion $\sigma$, which is the scale of smoothing. All the structures 
of the field below the scale $\sigma$ are smoothed out. It is worth to stress that the smoothing is over the 
fixed physical distance. Therefore we have introduced the factor $a^2$ in formula (\ref{Gauss}).
The scale of smoothing can be related either with the properties of 
the field or with the properties of the space. Namely, it can be for example a scale where the structure of a 
field becomes polymerised, as in the case studied in \cite{Hossain:2009ru, Hossain:2009vd}. Then effective 
field theoretical description is valid only at the scales larger than $\sigma$. Another source of 
smoothing can be the discrete nature of space. In this case averaging has to be performed on the scales where
the quantum gravitational effects becomes significant, namely on the Planck scale. Subsequent possible 
source of smoothing is decoherence of the quantum fluctuations \cite{Kiefer:2008ku,Sudarsky:2009qa}. 
In this case the parameter $\sigma$ encodes the details of interaction between the quantum fluctuations 
and the environment. Process of diffusion is also the natural smoothing factor.

In this paper we study how the effect of smoothing modify the power spectrum of the scalar field.
In our consideration we apply the smoothing to the quantum fluctuations generated during the phase
of inflation. In particular we show how power spectrum from de Sitter inflation is regularised 
by the smoothing. Then we investigate possible observational consequences of this effect. 

Now we will derive the spectrum of the scalar field in the flat FRW background in presence of smoothing. 
We will consider classical scalar field now, but the results obtained are the same like in the quantum case. 
The quantum case is discussed in more details in Appendix where also an example of the smoothed free scalar 
field in the Minkowski background is shown. 

The scalar field $\phi$ can be smoothed by performing the integration
\begin{equation}
\phi_{\Sigma}({\bf x},t) = \int_{V} d^3{\bf y}  \sqrt{q} \phi({\bf y},t) \Sigma({\bf x-y})  \label{conv}
\end{equation}
where $\Sigma$ is the shooting function. Taking the Fourier transform of equation (\ref{conv}) we obtain 
\begin{equation}
\phi_{\Sigma}({\bf k},t) = (2\pi)^{3/2} a^3 \phi({\bf k},t) \Sigma({\bf k}) \label{phif}
\end{equation}
where the Fourier transform of the smoothing function (\ref{Gauss}) is given by
\begin{equation}
\Sigma({\bf k}) \equiv \int \frac{d^3{\bf x}}{(2\pi)^{3/2}} \Sigma({\bf x}) e^{-i{\bf k \cdot x }} =
 \frac{1}{(2\pi)^{3/2}} \frac{1}{a^3} e^{-\frac{k^2\sigma^2}{2a^2}} 
\end{equation}
and where  $k^2={\bf k \cdot k}$. Squaring the equation (\ref{phif}) and taking the 
Fourier transform of the smoothing function, we find 
\begin{equation}
\left|\phi_{\Sigma}({\bf k},t)\right|^2 = (2\pi)^{3} a^6\left|\phi({\bf k},t)\right|^2 \left|\Sigma({\bf k})\right|^2
=\left|\phi({\bf k},t)\right|^2 e^{-\frac{k^2\sigma^2}{a^2}}. \label{phif2}
\end{equation}
The power spectrum of the field $\phi({\bf x},t)$ is defined as follows
\begin{equation}
\mathcal{P}(k)=\frac{k^3}{2\pi^2}  \left|\phi({\bf k},t)\right|^2. \label{power}
\end{equation}
The spectrum of the smoothed field $\phi_{\Sigma}({\bf x},t)$ can be expressed taking the (\ref{phif2}). Based 
on the definition (\ref{power}) we obtain
\begin{equation}
\mathcal{P}_{\Sigma}(k)=\mathcal{P}(k) e^{-\frac{k^2\sigma^2}{a^2}}. \label{PSigma}
\end{equation}
The effect of the smoothing is exponential suppression of the power spectrum on the scales $\lambda=a/k$ 
comparable with $\sigma$. In the next section we show how this effect modify the inflationary spectrum of 
quantum fluctuations.

\section{Regularisation of the inflationary spectrum}

\begin{figure}[ht!]
\centering
\includegraphics[width=10cm,angle=0]{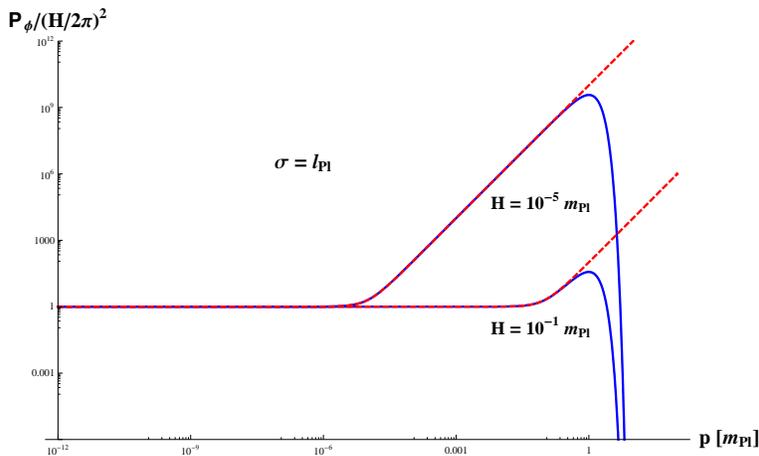}
\caption{Normalised power spectrum of the smoothed free scalar field generated during the de Sitter phase.
The scale of smoothing is fixed to be the Planck length. Dashed lines represents the case with no smoothing.}
\label{Pf}
\end{figure}

In this section we will consider quantum fluctuations of the scalar field in de Sitter inflation. 
During this phase of evolution the scale factor is given by 
\begin{equation}
a = -\frac{1}{H\eta},  
\end{equation}
where $H$ is Hubble factor which is constant. Parameter $\eta$ is conformal time defined as $\eta=\int\frac{dt}{a} \leq 0$. 
The vacuum is given by the Bunch-Davies mode 
\begin{equation}
f_k = \frac{e^{-ik\eta}}{\sqrt{2k}}\left(1+\frac{i}{k\eta}\right). 
\end{equation}
Applying this to the definition (\ref{Powerf}) and introducing the smoothing as in equation (\ref{PSigma}) we find
\begin{equation}
\mathcal{P}_{\phi} = \left(\frac{H}{2\pi}\right)^2\left[1+\left(\frac{p}{H}\right)^2 \right] e^{-p^2\sigma^2}.  \label{PfdS1}
\end{equation}
Here we have introduced physical momentum $p = \frac{k}{a}$ in order to simplify the notation. In Fig. \ref{Pf} we plot 
function (\ref{PfdS1}) for two values of $H$. We also assumed that $\sigma=l_{\text{Pl}}$. 
In the super-horizontal limit $p\rightarrow 0$ this spectrum takes the classical form
\begin{equation}
\mathcal{P}_{\phi} = \left(\frac{H}{2\pi}\right)^2. \label{PdS}
\end{equation}
However for the sub-horizontal modes the spectrum is significantly modified by the smoothing effect.
In the case with no smoothing, the spectrum is approximated, in this range, by the Minkowski vacuum 
and therefore $\mathcal{P}_{\phi}  \propto k^2$.  Such a behaviour leads to the quadratic UV divergence
of the correlation function. However due to the smoothing, the spectrum is suppressed and UV divergence
is removed. The transition between the Minkowski vacuum range and the smoothing domain is in the form 
of the bump in the power spectrum. Maximum of this bump is located in 
\begin{equation}
p_{\text{max}} = \frac{1}{\sigma}\sqrt{1-H^2\sigma^2}.
\end{equation}
However while  $H^2\sigma^2<1$ then the maximum of the spectrum is placed in $p=0$. In this case 
bump is not present in the power spectrum. The spectrum is not scale-invariant in this bump range.
It is namely "blue" on the left side with the spectral index $n_s=2$. Moreover amplitude of perturbations 
here can be much bigger that in the super-horizontal domain. This is interesting feature because e.g. models 
of primordial black holes formation require such a "blue" tilted spectrum in order to resolve problem of
the supermassive black holes \cite{Duechting:2004dk}. 

The spectrum obtained here is free from the UV divergences. However the IR (logarithmic) divergence remains
and can be removed only by taking superposition of the positive and negative Bunch-Davies modes 
(See e.g.\cite{Ford:1997hb}). This is naturally obtained e.g. in the models with the contracting (bounce) 
phase before the inflation (See \cite{Mielczarek:2009zw}). In this case the inflationary spectrum is suppressed 
also on the large scales (small $p$) and IR divergence is removed. If the lack of divergences is condition for 
the physical relevance then inflationary model with the bounce and smoothing is favoured. These both effects can
be a result of quantum nature of gravity.

It is interesting to see whether the studied effect of smoothing can lead to any observational imprints. 
In order to investigate it we will calculate observed primordial power spectrum. For the particular $k$ mode, 
the amplitude of perturbations can be calculated during the horizon crossing, namely for $k=aH$. This gives us
\begin{equation}
\mathcal{P}_{\phi}\vert_{k=aH} = \frac{H^2}{2\pi^2}e^{-H^2\sigma^2}.  \label{PfdS2}
\end{equation}
However this is very idealised case. In real, the modes become \emph{frozen} for $k \ll aH$. In this case spectrum is 
given by (\ref{PdS}) and any effects of smoothing are seen. Namely the smoothing factor is now $e^{-p^2\sigma^2}$
where $p^2\sigma^2 \ll H^2 \sigma^2$. If $H\approx 10^{-5}m_{\text{Pl}}$ as expected for the standard inflation and 
$\sigma \approx l_{\text{Pl}}$ then $p^2\sigma^2 \ll 10^{-10}$. Therefore the exponential factor is with very high 
precision equal to one. Therefore we conclude that the effects of smoothing from the inflation are unable to detect. 
This can be seen as no-go result, this way of searching for the quantum gravity effects is rather inaccessible.

\section{Smoothing and its possible effects on CMB} \label{sec:CMB}

In the previous section we have shown that in the case of de Sitter inflation, the 
effect of smoothing can be suppression of the total power spectrum. However 
when $H \ll 1/\sigma$ then this effect is negligible and cannot be studied observationally.
In this section we will check whether effects of suppression in any other form can 
be probed observationally. In particular the \emph{instantaneous smoothing} can lead to more 
significant modifications.  In this case smoothing is effective only in some short period of time,
therefore the smoothing factor can be written as $e^{-\frac{k^2\sigma^2}{a_*^2}}$ where $a_*$ 
is the value of the scale factor during the smoothing take place. Based on the CMB observations 
we can only constraint the joined factor $\frac{\sigma^2}{a_*^2}$. The example we will examine 
in details in this section is direct smoothing of the primordial perturbations. We will 
study here rather phenomenological form of modifications of the power spectrum and results obtained
can be later related with some specific model. In this case modification will be given by the 
factor $e^{-k^2\sigma^2}$.

For the perturbations generated during the phase of inflation the spectrum can be parametrised 
in the following form
\begin{eqnarray}
\mathcal{P}_s(k) &=& A_s \left( \frac{k}{k_0}\right)^{n_s-1}. \label{Ps}
\end{eqnarray} 
Introducing directly effect of smoothing we obtain the spectrum 
\begin{equation}
\mathcal{P}_{\Sigma}(k)=A_s \left( \frac{k}{k_0}\right)^{n_s-1} e^{-k^2\sigma^2}.  \label{Pphen}
\end{equation}
We have used here equation (\ref{PSigma}) with $a=1$. Based on this spectrum we compute spectrum 
of the CMB anisotropy and polarisation. We also find confidence intervals for the parameters of the 
model, namely on $A_s$, $n_s$ and $\sigma$. In the numerical calculations we use the publicly available 
CAMB \cite{Lewis:1999bs} code and Markov Chain Monte Carlo (MCMC) package Cosmo MC \cite{Lewis:2002ah}.
The codes were suitably modified to investigate the spectrum (\ref{Pphen}).  
To test our model we use the current astronomical observations from WMAP5  \cite{Komatsu:2008hk}. We 
take the standard cosmological parameters as follows
\begin{equation}
(H_0,\Omega_bh^2,\Omega_ch^2,\tau) = (70,0.022,0.12,0.09) 
\end{equation}
and the pivot scale $k_0=0.05\ \text{Mpc}^{-1}$.

\begin{figure}[ht!]
\centering
\begin{tabular}{c}
a)\includegraphics[width=6cm,angle=270]{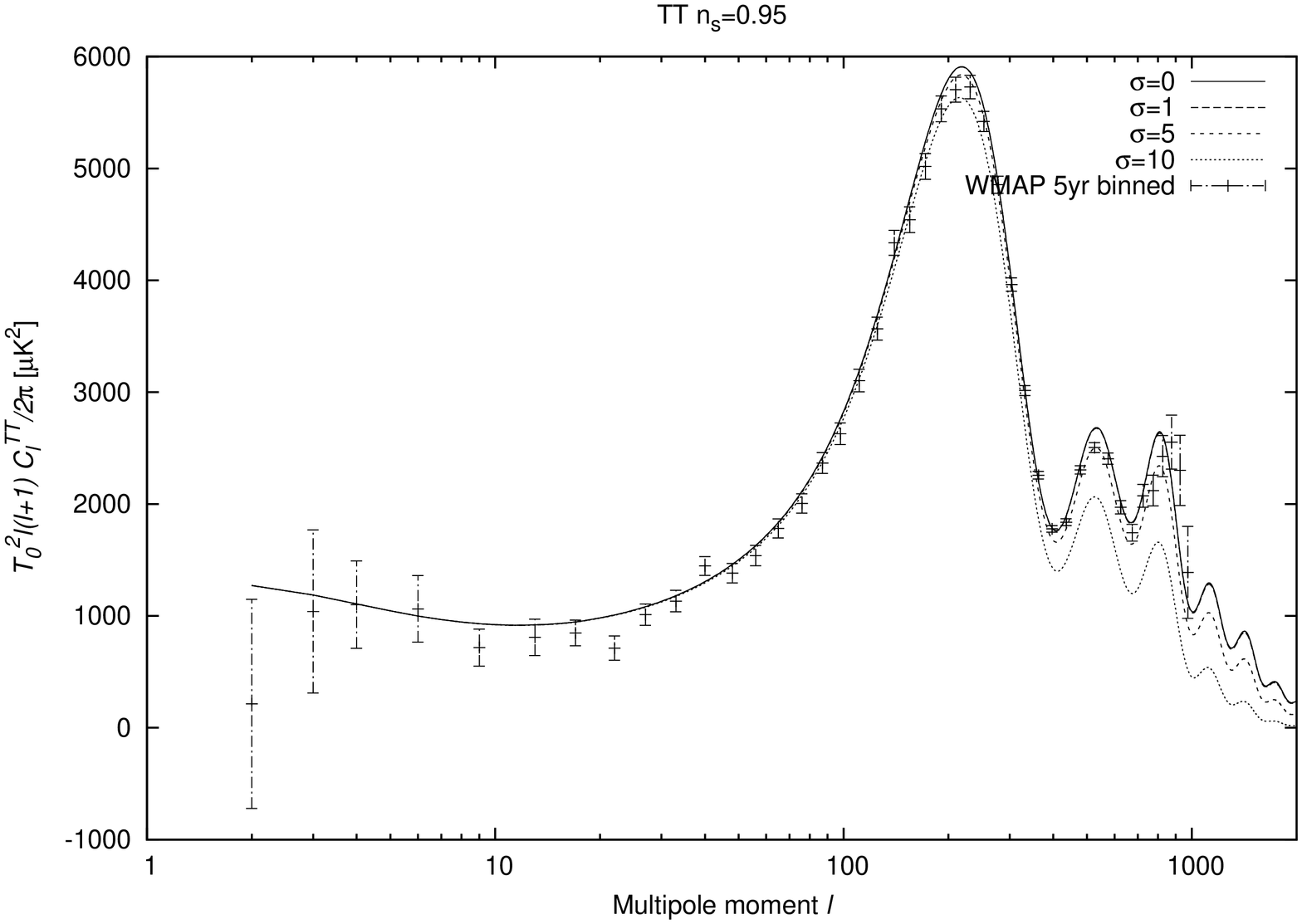}\\
b)\includegraphics[width=6cm,angle=270]{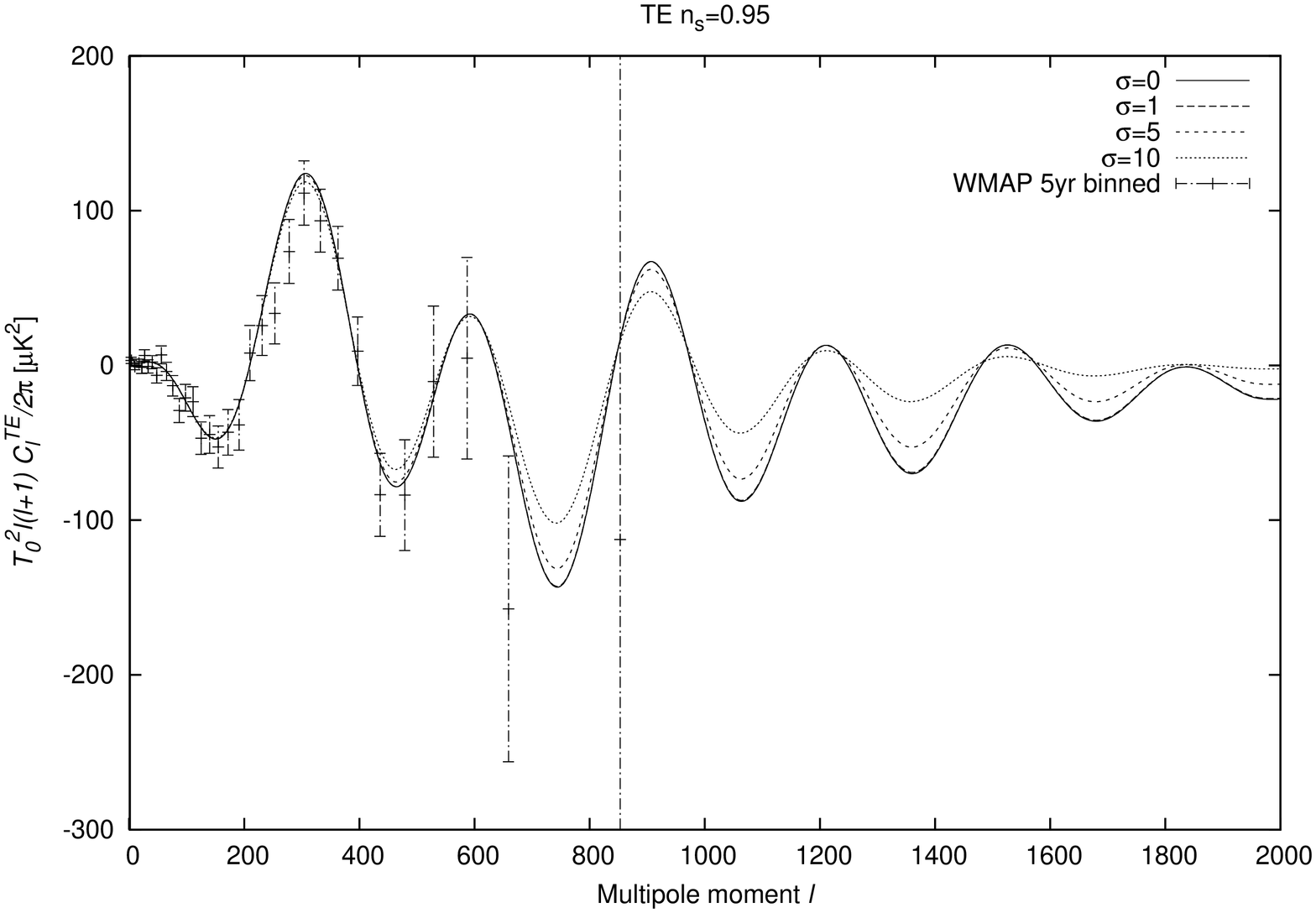}\\ 
c)\includegraphics[width=6cm,angle=270]{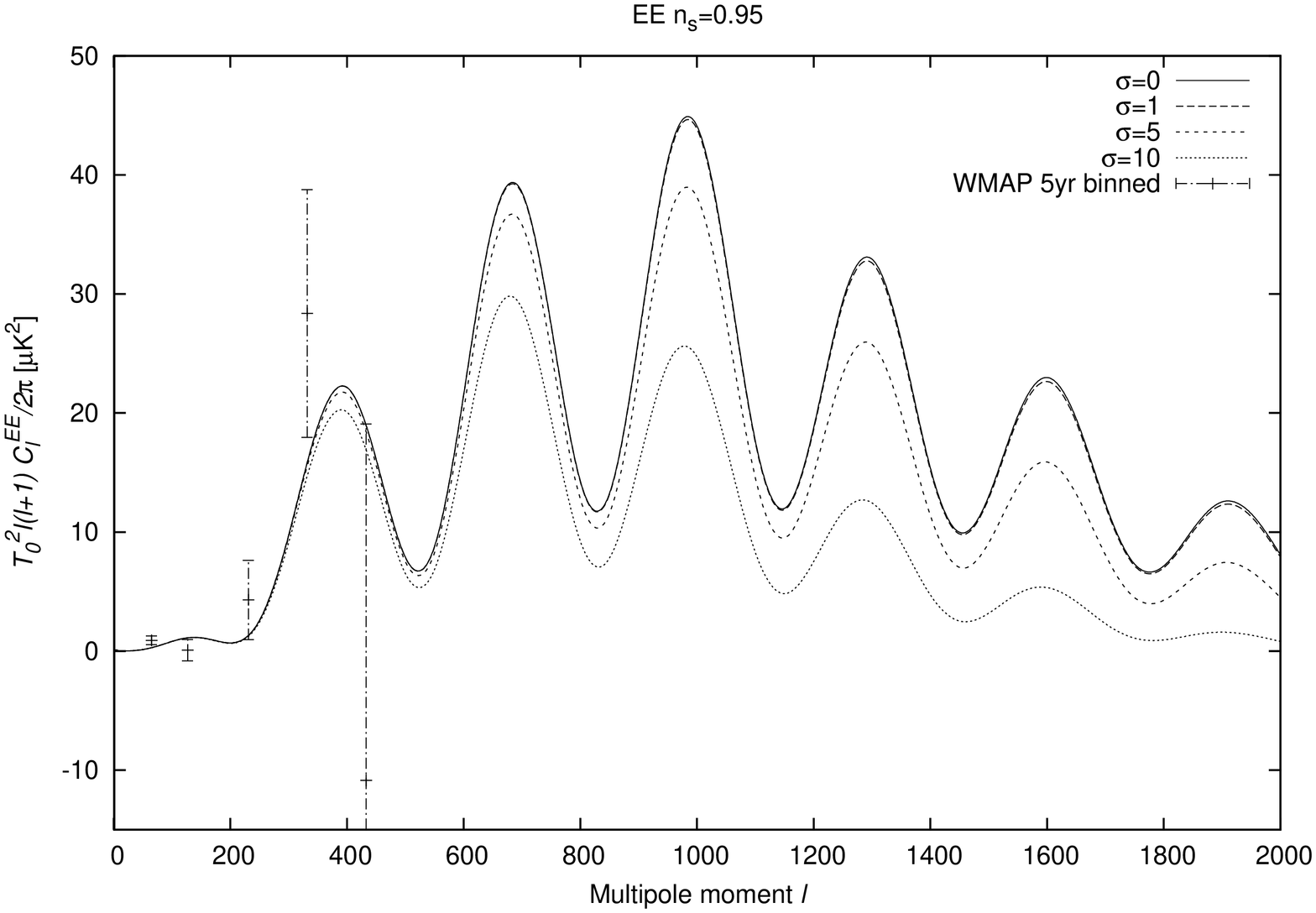}\\
\end{tabular}
\caption{Power spectra in functions of variation in $\sigma$: a) Temperature, b) Temperature-Polarisation, 
c) Polarisation. Here we have assumed $A_s=2.3\cdot 10^{-9}$.}
\label{CMB}
\end{figure}

In Fig. \ref{CMB} we show the obtained TT, TE and EE spectra of CMB. 
We show how these spectra are modified by the presence of smoothing. The evident signature is suppression of
the spectra on the high multipoles. This effect is opposite to the effect of low multipoles suppression, usually
considered in context of searching for the departures from the standard cosmological model. In particular 
the effect of the bounce can lead to suppression of low multipoles \cite{Mielczarek:2009zw}. However for the 
low multipoles, the so-called \emph{cosmic variance} becomes significant. This is serious obstacle in testing 
cosmological models beyond the standard inflation. On the high multipoles, the effect of \emph{cosmic variance}
becomes negligible. However instrumental effects becomes significant here. Resolution of observations can be 
nonetheless considerably improved by the new missions like Planck \cite{:2006uk}. 

\begin{figure}[ht!]
\centering
\includegraphics[width=12cm,angle=0]{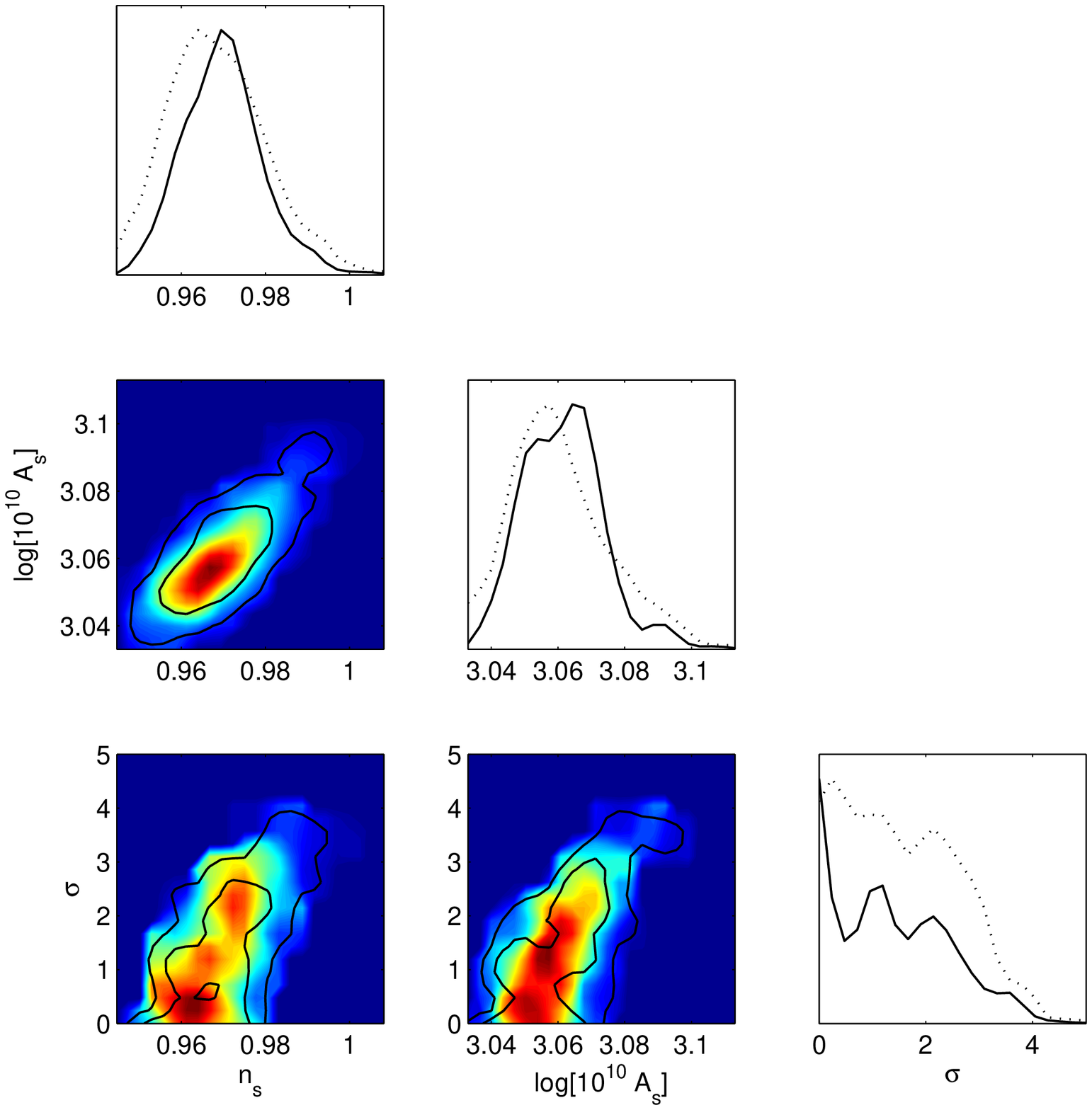}
\caption{Constraints for the parameters $A_s$, $n_s$ and $\sigma$. 2D plots: solid lines show the 68 and 95 $\%$ confidence intervals. 1D plots: dotted lines are mean likelihoods of samples, solid lines are marginalised probabilities.}
\label{Constraints}
\end{figure}

In Fig. \ref{Constraints} we show the obtained constraints on the parameters $A_s$, $n_s$ and $\sigma$. 
We see that values of parameters $n_s$ and $A_s$ are constrained from the both sides. In case of the parameter 
$\sigma$ only upper bound can be found. Based on the five years observations of WMAP satellite we find
\begin{equation}
\sigma <  \sigma_{\text{c}}=2.86 \ \text{Mpc (95\% CL)}.
\end{equation}
This constraint can be significantly (by a few orders of magnitude) improved with the future 
observations e.g. from Planck. The scale $\sigma_c$ corresponds to the present intergalactic distances. One can apply the 
obtained limit to the case of \emph{instantaneous smoothing}, then $\sigma \rightarrow \frac{\sigma}{a_*}$.
Let us study a case when the smoothing takes place at the end of inflation, it can be e.g. due to decoherence. 
Taking the $T_{\text{GUT}}\simeq 10^{14}$ GeV to be the energy scale at the end of inflation we find $a_* \simeq 10^{-27}$.
Based on this we find that at the end of inflation, the scale of smoothing is constrained by
\begin{equation}
\sigma < 100 \ \mu\text{m}.  
\end{equation}
This is far from the Planck scale, however relatively close to the scale of the micro-world where 
the quantum effects becomes significant. Therefore we speculate that the method can be used to constraint 
some models of decoherence of the quantum fluctuations. Also the scale of diffusion can be constrained
with this result.

\section{Summary} \label{sec:sum}

In this paper we have considered effects of the scalar field smoothing on the power spectrum.
We considered model where averaging function is given by Gaussian shape. Based on this we have
showed that smoothing modifies power spectrum by the factor $e^{-\frac{k^2\sigma^2}{a^2}}$, 
where $\sigma$ is dispersion of the Gaussian distribution. We have applied the smoothing 
to the quantum fluctuations generated during the phase of cosmic inflation. We have shown that 
the spectrum is regularised by the effect of smoothing and UV divergence is removed. Subsequently 
we have investigated whether smoothing on the Planck scale can be probed observationally. This
would be possible way to study quantum gravitational effects. Unfortunately, our investigations 
suggest rather no-go perspective for this method.  

Later we have investigated whether smoothing in any other form can be studied observationally.
We have considered phenomenological modification of the inflationary power spectrum due to 
the smoothing. We have shown that smoothing can lead to suppression of the high multipoles in the 
spectrum of the CMB anisotropy and polarisation. Based on five years data from WMAP satellite 
we have derived constraint for the parameter $\sigma$. We showed that $\sigma < \sigma_{\text{c}}=2.86$ 
Mpc (95$\%$ CL). This size is comparable with distances between galaxies in the galactic cluster.
We have applied this constraint to the model of \emph{instantaneous smoothing} at the end of inflation.
We showed that the constraint $\sigma < 100 \ \mu$m is satisfied then. This is far from the Planck scale
and cannot be used to probe quantum gravity effects. However models of other physical processes like 
diffusion or decoherence can be, in principle, constrained.

\appendix 
\section{Smoothed quantum scalar field} \label{Appendix1}

The quantum scalar field in FRW background can be decomposed for the Fourier modes as follows
\begin{equation}
\hat{\phi}({\bf x},\eta) = \frac{1}{a} \int \frac{d^3 {\bf k}}{(2\pi)^{3/2}}\left( f_k \hat{a}_{\bf k}  e^{i {\bf k \cdot x}}
+  f_k^* \hat{a}_{-{\bf k}}^{\dagger}  e^{-i {\bf k \cdot x}} \right).
\end{equation}
Here $f_k$ and $f_k^*$ are the so-called mode functions. Moreover creation and annihilation operator fulfils relations 
$[\hat{a}_{\bf k},\hat{a}^{\dagger}_{\bf p}]=\delta^{(3)}({\bf k-p})$ and  $[\hat{a}_{\bf k},\hat{a}_{\bf p}]=
[\hat{a}^{\dagger}_{\bf k},\hat{a}^{\dagger}_{\bf p}] =0$.
The averaging of the quantum field $\hat{\phi}({\bf x},\eta)$ is performed in the same way like in the classical case, namely
\begin{equation}
\hat{\phi}_{\Sigma}({\bf x},\eta) = \int_{V} d^3{\bf y} \sqrt{q} \hat{\phi}({\bf y},\eta) \Sigma({\bf x-y}). 
\end{equation}
Now it is straightforward to calculate the correlation function. We obtain 
\begin{equation}
\langle 0|\hat{\phi}_{\Sigma}({\bf x},\eta)\hat{\phi}_{\Sigma}({\bf y},\eta) |0 \rangle = 
\int_0^{\infty} \frac{dk}{k} \mathcal{P}_{\phi}(k) e^{-\frac{k^2\sigma^2}{a^2}} \frac{\sin(kr)}{kr} \label{Cor1}
\end{equation}
where the power spectrum is defined as follows
\begin{equation}
\mathcal{P}_{\phi}(k) = \frac{k^3}{2\pi^2} \frac{|f_k|^2}{a^2}. \label{Powerf}
\end{equation}
Absorbing the factor $e^{-\frac{k^2\sigma^2}{a^2}}$ in the integral (\ref{Cor1}) we can define the 
smoothed spectrum as in equation (\ref{PSigma}). Therefore the results obtained here as in the classical 
case are the same.

As an example we consider free scalar field in the Minkowski space. In this case $a=1$, $\eta=t$ and $f_k=\frac{e^{-ikt}}{\sqrt{2k}}$. 
The correlation function for the field without the smoothing is equal to 
\begin{equation}
\langle 0|\hat{\phi}({\bf x},t)\hat{\phi}({\bf y},t) |0 \rangle = \frac{1}{4\pi^2 r} \int_0^{\infty} dk \sin(kr) =\frac{1}{4\pi^2} \frac{1}{({\bf x-y})^2}.
\end{equation}
This is exactly the function (\ref{CM}). The expression was calculated introducing 
regularizator $e^{-\epsilon kr}$ and taking $\epsilon \rightarrow 0$
after the integration. This correlation function is UV divergent in the limit $|{\bf x-y}|\rightarrow 0$.
The correlator of the smoothed scalar field does not exhibit such a divergence. Namely it can expressed as 
follows
\begin{equation}
\langle 0|\hat{\phi}_{\Sigma}({\bf x},t)\hat{\phi}_{\Sigma}({\bf y},t) |0 \rangle =\frac{1}{4\pi^2\sigma^2} 
e^{-\frac{r^2}{4\sigma^2}} \frac{\sqrt{\pi}}{4}
\frac{\text{Erfi}\left(\frac{r}{2\sigma}\right)}{\left(\frac{r}{2\sigma}\right)} 
\end{equation}
where $r =|{\bf x-y}|$ and $\text{Erfi}(z)$ is imaginary error function defined 
as $\text{Erfi}(z)=\text{erf}(iz)/i$. We can check the UV limit of this function.
Using the limit for the imaginary error function 
\begin{equation}
\lim_{x\rightarrow 0} \frac{\text{Erfi}(x)}{x} =\frac{2}{\sqrt{\pi}} 
\end{equation}
we find 
\begin{equation}
\langle 0|\hat{\phi}_{\Sigma}(0,t)\hat{\phi}_{\Sigma}(0,t) |0 \rangle =\frac{1}{8\pi^2} \frac{1}{\sigma^2}.
\end{equation}
This value is finite and therefore UV divergence does not occur.


\begin{thebibliography}{99}

%\cite{Hossain:2009ru}
\bibitem{Hossain:2009ru}
  G.~M.~Hossain, V.~Husain and S.~S.~Seahra,
  %``Non-singular inflationary universe from polymer matter,''
  arXiv:0906.2798 [astro-ph.CO].
  %%CITATION = ARXIV:0906.2798;%%

%\cite{Hossain:2009vd}
\bibitem{Hossain:2009vd}
  G.~M.~Hossain, V.~Husain and S.~S.~Seahra,
  %``Background independent quantization and wave propagation,''
  arXiv:0906.4046 [hep-th].
  %%CITATION = ARXIV:0906.4046;%%

%\cite{Kiefer:2008ku}
\bibitem{Kiefer:2008ku}
  C.~Kiefer and D.~Polarski,
  %``Why do cosmological perturbations look classical to us?,''
  arXiv:0810.0087 [astro-ph].
  %%CITATION = ARXIV:0810.0087;%%

%\cite{Sudarsky:2009qa}
\bibitem{Sudarsky:2009qa}
  D.~Sudarsky and A.~De Unanue,
  %``A window to quantum gravity phenomena in the emergence of the seeds of
  %cosmic structure,''
  J.\ Phys.\ Conf.\ Ser.\  {\bf 174} (2009) 012059
  [arXiv:0901.2884 [gr-qc]].
  %%CITATION = 00462,174,012059;%%

%\cite{Duechting:2004dk}
\bibitem{Duechting:2004dk}
  N.~Duechting,
  %``Supermassive black holes from primordial black hole seeds,''
  Phys.\ Rev.\  D {\bf 70} (2004) 064015
  [arXiv:astro-ph/0406260].
  %%CITATION = PHRVA,D70,064015;%%

%\cite{Ford:1997hb}
\bibitem{Ford:1997hb}
  L.~H.~Ford,
  %``Quantum field theory in curved spacetime,''
  arXiv:gr-qc/9707062.
  %%CITATION = GR-QC/9707062;%%

%\cite{Mielczarek:2009zw}
\bibitem{Mielczarek:2009zw}
  J.~Mielczarek,
  %``The Observational Implications of Loop Quantum Cosmology,''
  arXiv:0908.4329 [gr-qc].
  %%CITATION = ARXIV:0908.4329;%%

%\cite{Lewis:1999bs}
\bibitem{Lewis:1999bs}
  A.~Lewis, A.~Challinor and A.~Lasenby,
  %``Efficient Computation of CMB anisotropies in closed FRW models,''
  Astrophys.\ J.\  {\bf 538} (2000) 473
  [arXiv:astro-ph/9911177].
  %%CITATION = ASJOA,538,473;%%

%\cite{Lewis:2002ah}
\bibitem{Lewis:2002ah}
  A.~Lewis and S.~Bridle,
  %``Cosmological parameters from CMB and other data: a Monte-Carlo approach,''
  Phys.\ Rev.\  D {\bf 66} (2002) 103511
  [arXiv:astro-ph/0205436].
  %%CITATION = PHRVA,D66,103511;%%

%\cite{Komatsu:2008hk}
\bibitem{Komatsu:2008hk}
  E.~Komatsu {\it et al.}  [WMAP Collaboration],
  %``Five-Year Wilkinson Microwave Anisotropy Probe (WMAP\altaffilmark 1 )
  %Observations:Cosmological Interpretation,''
  Astrophys.\ J.\ Suppl.\  {\bf 180} (2009) 330
  [arXiv:0803.0547 [astro-ph]].
  %%CITATION = APJSA,180,330;%%

%\cite{:2006uk}
\bibitem{:2006uk}
    [Planck Collaboration],
  %``Planck: The scientific programme,''
  arXiv:astro-ph/0604069.
  %%CITATION = ASTRO-PH/0604069;%%

\end{thebibliography}
\end{document}